\documentclass[fleqn,usenatbib,letters]{mnras}

\usepackage[T1]{fontenc}

\DeclareRobustCommand{\VAN}[3]{#2}
\let\VANthebibliography\thebibliography
\def\thebibliography{\DeclareRobustCommand{\VAN}[3]{##3}\VANthebibliography}

\usepackage{graphicx}	
\usepackage{amsmath}	
\usepackage{amssymb}	
\usepackage{multirow}
\usepackage{siunitx}
\usepackage{threeparttable}
\usepackage{cleveref}
\usepackage{pdflscape}
\usepackage{lscape}

\DeclareSIUnit \parsec 	    {pc}
\DeclareSIUnit \eV 			{eV}
\DeclareSIUnit \keV 		{keV}
\DeclareSIUnit \Msun 		{M_\odot}
\DeclareSIUnit \year        {yr}

\title[Hinting a DM nature of Sgr~A* using the S-stars]{Hinting a dark matter nature of Sgr~A* via the S-stars}

\author[E.~A.~Becerra-Vergara et al.]{
E.~A.~Becerra-Vergara,$^{1,2,3}$
C.R.~Arg\"uelles,$^{1,2,4}$
A.~Krut,$^{1,2}$
J.~A.~Rueda$^{1,2,5,6,7}$\thanks{E-mail: jorge.rueda@icra.it (JAR)}
and R.~Ruffini$^{1,2,5,6}$
\\
$^{1}$ICRANet, Piazza della Repubblica 10, 65122 Pescara, Italy\\
$^{2}$ICRA, Dip. di Fisica, Sapienza Universit\`a di Roma, P.le Aldo Moro 5, 00185 Rome, Italy\\
$^{3}$\textit{GIRG}, Escuela de F\'isica, Universidad Industrial de Santander, 680002 Bucaramanga, Colombia\\
$^{4}$Fac. de Ciencias Astron. y Geof\'isicas, Universidad Nacional de La Plata, Paseo del Bosque, B1900FWA La Plata, Argentina\\
$^{5}$ICRANet-Ferrara, Dip. di Fisica e Scienze della Terra, Universit\`a degli Studi di Ferrara, Via Saragat 1, 44122 Ferrara, Italy\\
$^{6}$INAF, Istituto de Astrofisica e Planetologia Spaziali, Via Fosso del Cavaliere 100, 00133 Rome, Italy
}

\date{Accepted XXX. Received YYY; in original form ZZZ}

\pubyear{2021}

\begin{document}
\label{firstpage}
\pagerange{\pageref{firstpage}--\pageref{lastpage}}
\maketitle

\begin{abstract}
The motion data of the S-stars around the Galactic center gathered in the last \SI{28}{\year} imply that Sgr~A* hosts a supermassive compact object of about \SI{4E6}{\Msun}, a result awarded with the Nobel Prize in Physics 2020. A non-rotating black hole (BH) nature of Sgr~A* has been uncritically adopted since the S-star orbits agree with Schwarzschild geometry geodesics. The orbit of S2 has served as a test of General Relativity predictions such as the gravitational redshift and the relativistic precession. The central BH model is, however, challenged by the G2 post-peripassage motion and by the lack of observations on event-horizon-scale distances robustly pointing to its univocal presence. We have recently shown that the S2 and G2 astrometry data are better fitted by geodesics in the spacetime of a self-gravitating dark matter (DM) \textit{core -- halo} distribution of \SI{56}{\keV}-fermions, ``\textit{darkinos}'', which also explains the outer halo Galactic rotation curves. This Letter confirms and extends this conclusion using the astrometry data of the $17$ best-resolved S-stars, thereby strengthening the alternative nature of Sgr~A* as a dense core of darkinos.
\end{abstract}




\section{Introduction}\label{sec:1}

The gravitational potential in the Galactic Center (GC) is dominated by a supermassive compact object, Sagittarius A* (Sgr~A*), long thought to be a massive black hole BH of $\approx \SI{4E6}{\Msun}$ \citep{2005ApJ...620..744G, 2008ApJ...689.1044G, 2010RvMP...82.3121G,2018A&A...618L..10G}. From the observational viewpoint, this inference on the nature of Sgr~A* mainly comes from the nearly Keplerian orbits of tens of stars belonging to the S-star cluster \citep{2009ApJ...692.1075G, 2017ApJ...837...30G}, whose motions are well described by geodesics in the Schwarzschild spacetime geometry. The most important S-cluster member is S2 which, with an orbital period of about \SI{16}{\year} and a pericenter of about {$1500$} Schwarzschild radii, has the most-compact orbit around Sgr~A*. The S2 orbit data have allowed to test General Relativity predictions such as the relativistic redshift \citep[see, e.g.,][]{2018A&A...615L..15G, 2019Sci...365..664D} and precession {\citep[see, e.g.,][]{2017ApJ...845...22P,2020A&A...636L...5G}}. However, not every news is good for the BH model; it is challenged by the G2 motion which cannot be explained by any geodesics in the BH geometry \citep{2017ApJ...840...50P, 2019ApJ...871..126G}, as well as by very scarce data at event-horizon-scale distances from Sgr~A*, robustly pointing to a univocal central BH presence \citep[see, e.g.,][]{2014ARA&A..52..529Y, 2019ApJ...884..148B}.


In view of the above, we have dived into the possibility of an alternative nature for Sgr~A* based on the fermionic DM profile predicted by the Ruffini-Arg\"uelles-Rueda (RAR) model  \citep{2015MNRAS.451..622R, 2018PDU....21...82A}. In the RAR model, the DM distribution in galaxies is obtained from the general relativity field equations, assuming it as a self-gravitating system of fermions at finite temperature in equilibrium, and distributed in phase-space according to the Fermi-Dirac statistics including a particle energy cutoff which gives to the configuration a finite size \citep[see][for more details]{2018PDU....21...82A}. We hereafter refer to these neutral, massive DM fermions as ``\textit{darkinos}''. The RAR model leads to a \textit{dense~core -- diluted~halo} density profile in which the darkinos are: 1) in a quantum degenerate regime within the nearly uniform core, 2) followed by an intermediate quantum-classical regime in the density falloff and plateau phase, and 3) finally in a Boltzmann regime in the outer halo that follows a power-law density ending with a nearly exponential cutoff defining the galaxy border. There is a bunch of astrophysical consequences of the \textit{core -- halo} profile of darkinos derived from the RAR model. In \citet{2018PDU....21...82A}, it has been shown that it explains the rotation curves of the Milky Way outer halo. In \citet{2019PDU....24..278A}, this agreement has been shown to apply as well to other galaxy types ranging from dwarfs to big ellipticals and galaxy clusters. These results have further enticed attention on the darkinos microphysics, e.g. their self-interactions \citep{2016JCAP...04..038A, 2020PDU....3000699Y} and interaction with neutrinos \citep{2020EPJC...80..183P}; as well as in their macrophysics, e.g. their lensing properties \citep{2016PhRvD..94l3004G}, their influence in the dynamics of binaries \citep{2017PhRvD..96f3001G}, their halo formation and stability on cosmological timescales \citep{2020MNRAS.tmp.3770A}, and their role in the large and small scale structure formation \citep{2020JCAP...09..041Y}.
 
Having recalled the overall features of the darkinos of the RAR model, we turn now back to the topic of this Letter. We have shown in \citet{2020A&A...641A..34B} that, for darkinos of \SI{56}{\keV} rest mass-energy, the spacetime geometry produced by the dense quantum core leads to geodesics which fit equally good, and definitely superior, respectively, the observational data of S2 and G2. This result has given a first observational support to the darkinos alternative nature of Sgr~A*. Our aim here is to go a step further, and extend our previous analysis to the up-to-date astrometry data of the $17$ best-resolved S-stars \citep{2008ApJ...689.1044G, 2009ApJ...707L.114G, 2017ApJ...837...30G}. In this way, we are testing the models with a more robust sample composed of a statistically significant number of stars with well-determined positions and velocities. This considerably improves our previous test with S2 complemented by the object G2 which is of a questioned nature \citep[see, e.g.,][]{2014ApJ...796L...8W,2020Natur.577..337C}. We show below that the novel results here presented confirm and strengthen the alternative nature of Sgr~A* as a dense core of darkinos.

\section{Geodesics and astrometry data fit}\label{sec:2}


The monitoring of the S-stars around Sgr~A* provides crucial knowledge about the properties of the gravitational potential of the massive object hosted by Sgr~A*. One of the most interesting S-stars is S2, whose orbit determination is less prone to errors being it the brightest. It describes a nearly elliptical orbit with one of the shortest orbital periods ($\approx \SI{16}{\year}$; see e.g. \citealp{2003ApJ...586L.127G, 2017ApJ...837...30G, 2018A&A...615L..15G}), with its pericenter being the second closest to Sgr~A*, $r_{p(S2)}\approx \SI{0.6}{m\parsec}$. Therefore, S2 provides the most accurate constraints on the gravitational potential of Sgr~A* to date \citep{2009ApJ...707L.114G,2017ApJ...837...30G,2008ApJ...689.1044G}. 

\begin{figure}%
	\centering%
	\includegraphics[width=\hsize]{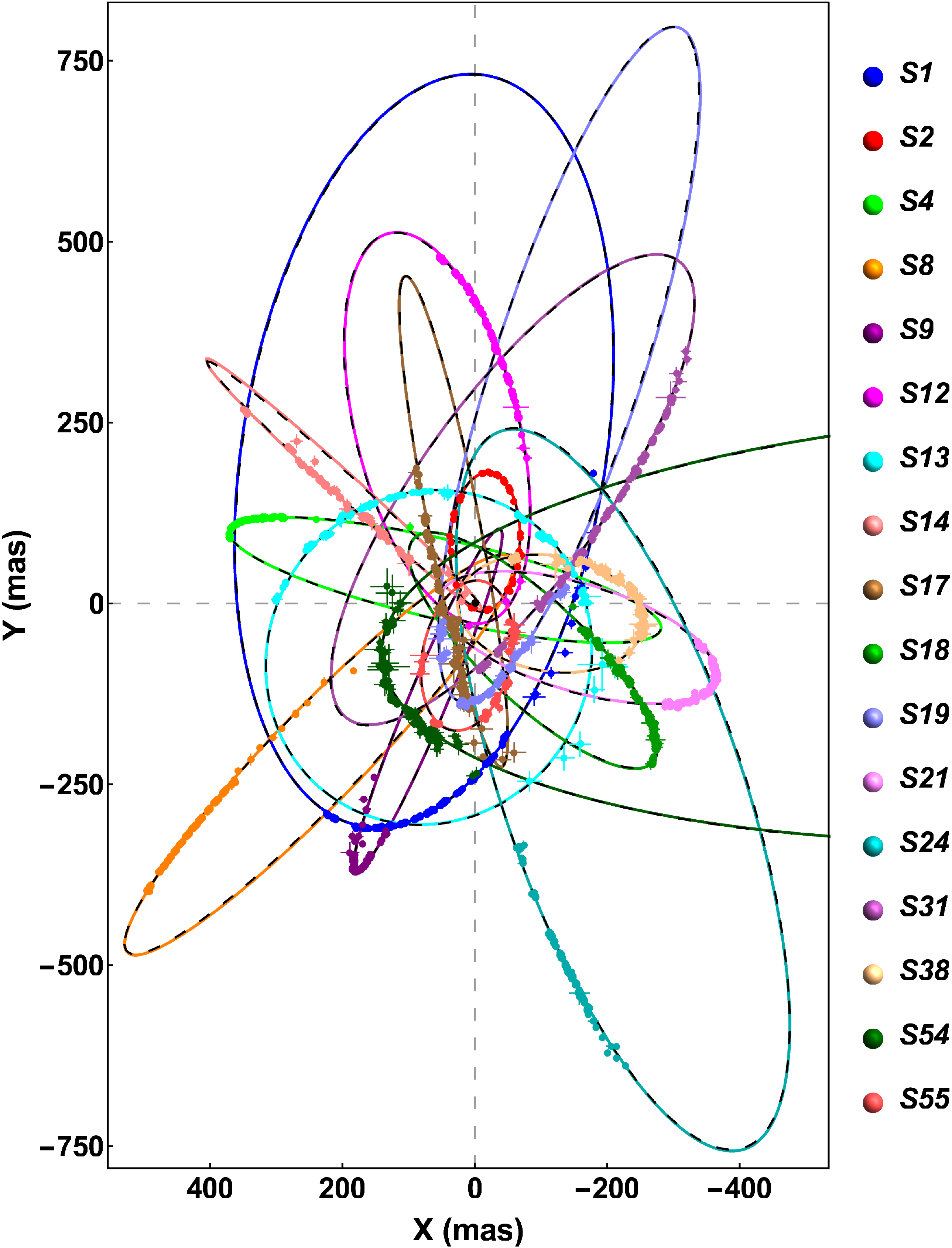}
	\caption{Best-fit orbits for the $17$ best-resolved S-star orbiting Sgr~A*. It shows the projected orbit on the sky, $X$ vs. $Y$, where $X$ is right ascension and $Y$ is declination. The \textit{black dashed curves} correspond to the BH model and the \textit{colored curves} to the RAR model of darkinos. We refer to \cref{tab:parameter} for the orbital parameters of each star in both models. The astrometric measurements are taken from \citet{2009ApJ...692.1075G,2017ApJ...837...30G,2019Sci...365..664D}.}
	\label{fig:Orbits}%
\end{figure}

We here follow the treatment described in Appendices A and C of \citet{2020A&A...641A..34B} for the best fitting procedure of S2 and G2, which has been used to constrain the model parameters in the two scenarios: the BH case and the (RAR) DM-core case. In the former, the relevant parameter associated with the source of the gravitational field is the BH mass ($M_{\rm BH}$), and in the latter, the DM core mass ($M_c$). The value of $M_c$ depends on the (underlying) free RAR-parameters, including the darkino mass $m$ \citep{2018PDU....21...82A}, but the reduced-$\chi^2$ minimization here applied to fit the orbits, following \citet{2020A&A...641A..34B}, only uses $M_c$. For each $M_c$ value applied in this procedure, the set of underlying free RAR model parameters is such that the overall \textit{core -- halo} RAR profile better fits the Galaxy rotation curve (see Appendix A  of \citealp{2020A&A...641A..34B} for further details). Then, we solve the full general relativistic equations of motion of a test particle in the set spacetime geometry and obtain the real geodesic which, projected onto the sky plane, best fit the apparent stellar orbit. At any given time, this is given by the measurements of the right ascension ($X$) and declination ($Y$). For a fixed model, the real orbit is determined once  values of the energy and angular momentum per unit mass of the geodesic are given. They can be determined using the effective potential technique as described in \citet{2020A&A...641A..34B}, by setting values for the pericenter and apocenter radii, $r_p$ and $r_a$. Alternatively, $r_p$ and $r_a$ could be replaced by the semi-major axis $a$ and eccentricity $e$ of an effective ellipse. The values of $r_p$ and $r_a$ are not affected in the projection of the orbit onto the sky plane, so the apparent orbit is then determined for given values of the osculating orbital elements, i.e. $\omega$, $i$, and $\Omega$, respectively, the argument of pericenter, the inclination between the real orbit and the sky plane, and the ascending node angle. With this, the orbital period $P$ and the time of closest approach to the GC, i.e. the time to reach the pericenter, $t_p$ (in J2000 time convention; see \citealp{2020A&A...641A..34B} for details), can be also inferred. Constant position offsets $X_0$ and $Y_0$ are also introduced to account for the relative position of the gravitational center of mass to the reference frame \citep[see Eq. C12 in][and references therein]{2020A&A...641A..34B}. The procedure is performed in an iterative fashion to obtain the best-fit parameters from least squares minimization. In \citet{2020A&A...641A..34B}, the application of this procedure to the case of S2 led to $M_{\rm BH} = \SI{4.075E6}{\Msun}$, in the BH model, and $M_c = \SI{3.5E6}{\Msun}$, in the RAR model. This $M_c$ value together with the overall rotation curve fit, implied the minimum allowed darkino mass, $m c^2=\SI{56}{\keV}$, fulfilling all observational constraints. For this mass, the DM core radius is $r_c\approx \SI{0.4}{\milli\parsec}$ \citep{2020A&A...641A..34B}. Larger darkino masses (up to $\SI{345}{\keV}$), for the same $M_c$, imply more compact DM core sizes down to a few Schwarzschild radii, still satisfying the rotation curve data \citep{2018PDU....21...82A}.

We emphasize the reliability of our fitting procedure. Our inferred value of $M_{\rm BH}$ in the BH case, using the S2 data, agrees with the most recently reported values, e.g. $M_{\rm BH} = \SI{4.1E6}{\Msun}$ by \citet {2018A&A...615L..15G}, and \SI{3.975E6}{\Msun} by \citet{2019Sci...365..664D}. We here extend the application of our model, previously tested with S2 and G2, to the other S-stars. We apply the above procedure keeping fixed the above parameters since they define the source of the gravitational field. Likewise, we fix $X^{\rm BH}_{0} = -0.0830$ and $Y^{\rm BH}_{0} = 2.4893$ (units of milliarcsecond), $X^{\rm RAR}_{0} = -0.1557$, $Y^{\rm RAR}_{0} = 2.5527$, and the distance to Sgr~A*, $\SI{8}{\kilo\parsec}$. We then search for the parameters that determine the real orbit and best fit the apparent one. We analyze the $17$ best-resolved S-stars S1, S2, S4, S8, S9, S12, S13, S14, S17, S18, S19, S21, S24, S31, S38, S54, and S55 \citep{2017ApJ...837...30G}.

\begin{figure*}%
	\centering%
	\includegraphics[width=0.95\hsize,clip]{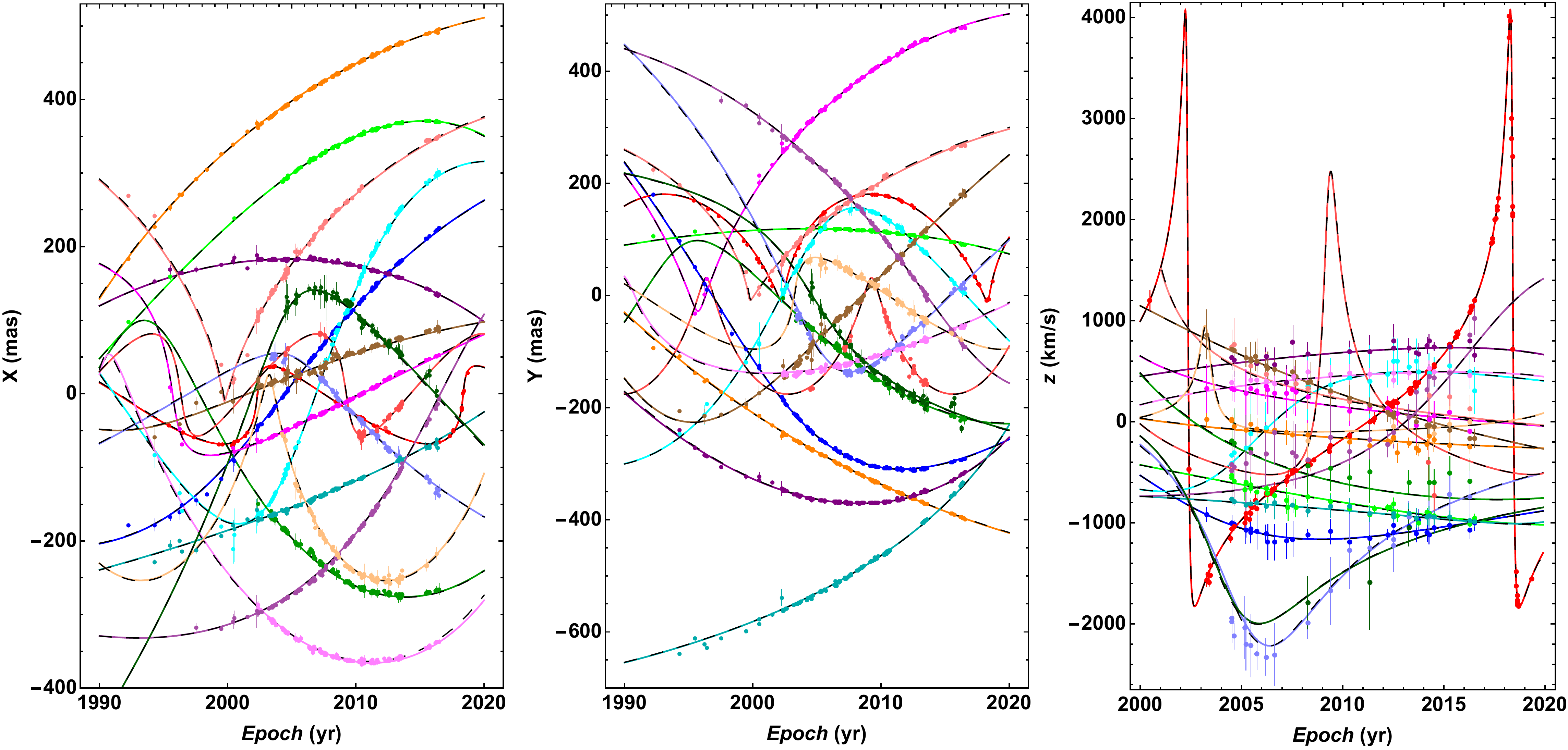}
	\caption{Best-fit of the observed right ascension $X$ (left panel), declination $Y$ (central panel), and line-of-sight radial velocity (redshift function) $z$ (right panel), as a function of time, for the $17$ best-resolved S-star orbiting Sgr~A*. The \textit{black dashed curves} correspond to the BH model and the \textit{colored curves} to the RAR model of darkinos. We refer to \cref{tab:parameter} for the orbital parameters of each star in both models. The astrometric measurements are taken from \citet{2009ApJ...692.1075G,2017ApJ...837...30G,2019Sci...365..664D}.}
	\label{fig:XYZvst}%
\end{figure*}

\section{Discussion and conclusions}\label{sec:3}

\Cref{tab:parameter} summarizes the best-fit model parameters and the corresponding reduced-$\chi^2$ for the position ($X$ and $Y$) and the line-of-sight radial velocity (i.e. the redshift function $z$), for the central BH and the RAR DM models. We can see that, overall, the RAR model performs slightly better than the BH model. An estimate of the performance can be obtained by calculating the average of the averages, namely to sum up the values of the last column of \cref{tab:parameter}, and divide it by the sample number. For the RAR model, this estimate leads to $1.5741$, and for the BH model, $1.6273$. This confirms their comparable accuracy in describing the S-stars data, being the RAR model of darkinos slightly preferred.

We can gain information on the reliability of our fitting procedure by comparing the inferred parameters in the BH case. We have recalled in \cref{sec:2} that our BH mass inference using S2 data agrees with most of the recent values reported in the literature using the same object. In addition, it also agrees with the reported value obtained from the simultaneous fit of several stellar orbits: \citet{2016ApJ...830...17B} reported $M_{\rm BH} = \SI{4.02 \pm 0.16E6}{\Msun}$ using S2 and S38, and \citet{2017ApJ...845...22P} reported $M_{\rm BH} = \SI{4.15 \pm 0.13 E6}{\Msun}$ using S2, S38, and S55. This is further confirmed by the fact that, our inferred orbital parameters for the BH model (see \cref{tab:parameter}), do not differ by more than \SI{2}{\percent} for S2, \SI{1}{\percent} for S38 and \SI{3}{\percent} for S55 from the values reported in \citet{2017ApJ...845...22P}. Our inferred value of $M_{\rm BH}$ is also within the window reported in \citet{2017ApJ...837...30G}, $M_{\rm BH} = \SI{4.28 \pm 0.21 E6}{\Msun}$, for the $17$ S-stars. While these estimates of $M_{\rm BH}$ in the existing literature used Post-Newtonian approximations, our method is fully general relativistic \citep[see][]{2020A&A...641A..34B}.

Having said this, we can turn to the visualization of the orbits. \Cref{fig:Orbits} shows the data and a comprehensive plot of the best-fit of the observed orbits of the sample of $17$ S-stars, including the stars with the most compact orbits (S2, S38, and S55). The similar performance of the RAR and BH models is evident, being their differences almost indistinguishable at these scales.
In \cref{fig:XYZvst}, we present the data and the best-fit of the time evolution of the star position components, $X(t)$ and $Y(t)$, as well as the redshift function, $z(t)$, for the S-star sample. Also in this case, the similar performance of the two tested models is evident. This is particularly relevant because that a model provides an accurate fit of the orbit, i.e. of $X$ vs. $Y$, does not necessarily imply that it correctly fits its time evolution, namely the model must correctly predict the observed star motion. Therefore, as described in \citet{2020A&A...641A..34B}, the estimation of the goodness of the fit must compare the theoretical values of $X$, $Y$ and $z$ with the measured values at each observational time.

Summarizing, this Letter shows for the first time that a highly dense concentration of DM particles sited at the GC can explain the dynamics of the S-stars with similar (and some cases better) accuracy compared to a central BH model. These results strengthen the alternative nature of Sgr~A* as a dense quantum core of darkinos superseding the central massive BH scenario. There is the key additional fact that this very same \textit{core -- halo} distribution of \SI{56}{\keV} darkinos also explains the rotation curves of the Milky Way \citep[see][for details]{2018PDU....21...82A,2020A&A...641A..34B}. For particle masses $\sim$ \SI{100}{\kilo\eV}, {the core radius shrinks from $\SI{0.4}{\milli\parsec}$ to a few Schwarzschild radii, so} the gravitational potential produced by a central BH of mass $M_{\rm BH}$ and a (RAR) DM core of mass $M_c = M_{\rm BH}$, practically coincide for $r\gtrsim 10~G M_{\rm BH}/c^2$ \citep{2016PhRvD..94l3004G}. Therefore, the dynamics of baryonic matter at these scales {should not differ much in} the two scenarios. This becomes relevant for the dynamics of objects in the vicinity of Sgr A*, e.g. the recently detected \textit{hot-spots} claimed to move in a circular orbit of $7$--$23~G M_{\rm BH}/c^2$ radius \citep{2018A&A...618L..10G, 2020MNRAS.497.2385M}. However, this wide range of values shows how the inferred real orbit is strongly affected by model assumptions and the relatively poor quality of the spots astrometry data, which is not comparable with the S-stars data here used. The dynamics of these spots remain an interesting target for future investigation as the quantity and quality of the data improve. In this line, the recent observations of a new set of S-stars (S62, S4711--S4714), possibly reaching pericenter distances $\sim 400~G M_{\rm BH}/c^2$ {\citep{2020ApJ...889...61P,2020ApJ...899...50P}}, could also offer the possibility to further constrain the DM {core size around Sgr~A*, likewise the lower limit of the darkino mass}.

We would like to outline some additional astrophysical and cosmological consequences of the \textit{core -- halo} distribution of darkinos in the RAR model. First, it has been shown in \citet{2019PDU....24..278A} that the DM RAR profiles are \textit{Universal}, thereby can be also successfully applied to dwarfs, ellipticals, and galaxy clusters, for $m \approx \SI{50}{\keV}$. Second, a crucial question that arises is whether or not a DM profile with this morphology can be formed in a cosmological framework. Importantly, it has been recently demonstrated in \citet{2020MNRAS.tmp.3770A} that such \textit{core -- halo} profiles are natural outcomes within non-linear structure formation in warm DM cosmologies, when the fermionic (quantum) nature of the DM particles is accounted for. It has been there shown that these novel DM profiles are either thermodynamically and dynamically stable for the lifetime of the Universe, or eventually collapse into a supermassive BH if a critical (threshold) mass of the quantum core is reached. This provides a new appealing scenario for the formation, starting from a DM seed, of the supermassive BHs observed in active galactic nuclei (AGN), with key implications for AGN astrophysics and early cosmology \citep{2020MNRAS.tmp.3770A}.

\section*{Acknowledgements}

We thank the Referee for the many constructive suggestions which have definitely improved our presentation. E.A.B-V. thanks financial and research support from COLCIENCIAS under the program Becas Doctorados Nacionales 727, the International Center for Relativistic Astrophysics Network (ICRANet), Universidad Industrial de Santander (UIS), and the International Relativistic Astrophysics Ph.D Program.


\section*{Data Availability}

The astrometric data used in this work were obtained from \citet{2009ApJ...692.1075G, 2017ApJ...837...30G, 2019Sci...365..664D}, and the data here generated are available in \Cref{tab:parameter}.



\begin{landscape}
\begin{table}
\begin{center}
\caption{Summary of the inferred best-fit parameters of the BH model and the RAR model of darkinos using the astrometry data of the $17$ best-resolved S-stars.}\label{tab:parameter}
\resizebox{1.33\textwidth}{!}{%
\begin{tabular}{ccccccccccccccccccccccccccccc}
\hline
Star &  & Model &  & $a$ (as) &  & $e$ &  & $r_{p}$ (as) &  & $r_{a}$ (as) &  & $i$ $(\si{\degree})$ &  & $\omega$ $(\si{\degree})$ &  & $\Omega$ $(\si{\degree})$ &  & $P$ (\si{\year}) &  & $t_{p}$ (\si{\year}) &  & $\bar{\chi}^2_X$ &  & $\bar{\chi}^2_Y$ &  & $\bar{\chi}^2_z$ &  & $\langle \bar{\chi}^2\rangle$ \\ \cline{1-1} \cline{3-3} \cline{5-5} \cline{7-7} \cline{9-9} \cline{11-11} \cline{13-13} \cline{15-15} \cline{17-17} \cline{19-19} \cline{21-21} \cline{23-23} \cline{25-25} \cline{27-27} \cline{29-29}
 &  &  &  &  &  &  &  &  &  &  &  &  &  &  &  &  &  &  &  &  &  &  &  &  &  &  &  & \\
\multirow{2}{*}{S1} &  & RAR &  & $0.5940$ &  & $0.5530$ &  & $0.2655$ &  & $0.9225$ &  & $119.48$ &  & $122.22$ &  & $342.34$ &  & $165.93$ &  & $2001.62$ &  & $1.3368$ &  & $1.6463$ &  & $0.2661$ &  & $1.0831$ \\
 &  & BH &  & $0.5937$ &  & $0.5533$ &  & $0.2652$ &  & $0.9222$ &  & $119.33$ &  & $122.23$ &  & $342.39$ &  & $165.66$ &  & $2001.63$ &  & $1.5804$ &  & $1.2103$ &  & $0.2803$ &  & $1.0237$ \\
\hline
\multirow{2}{*}{S2} &  & RAR &  & $0.1252$ &  & $0.8866$ &  & $0.0142$ &  & $0.2361$ &  & $134.35$ &  & $66.772$ &  & $228.02$ &  & $16.054$ &  & $2018.38$ &  & $1.5964$ &  & $6.3411$ &  & $1.5964$ &  & $3.0725$ \\
 &  & BH &  & $0.1252$ &  & $0.8863$ &  & $0.0143$ &  & $0.2362$ &  & $134.35$ &  & $66.450$ &  & $227.97$ &  & $16.051$ &  & $2018.38$ &  & $1.8004$ &  & $7.2332$ &  & $1.8004$ &  & $3.3586$ \\
\hline
\multirow{2}{*}{S4} &  & RAR &  & $0.3569$ &  & $0.3895$ &  & $0.2179$ &  & $0.4958$ &  & $80.942$ &  & $290.82$ &  & $258.82$ &  & $77.508$ &  & $2034.71$ &  & $1.4043$ &  & $2.0123$ &  & $0.9533$ &  & $1.4566$ \\
 &  & BH &  & $0.3568$ &  & $0.3891$ &  & $0.2180$ &  & $0.4956$ &  & $80.876$ &  & $291.02$ &  & $258.82$ &  & $77.184$ &  & $2034.58$ &  & $1.6329$ &  & $1.6882$ &  & $0.9530$ &  & $1.4247$ \\
\hline
\multirow{2}{*}{S8} &  & RAR &  & $0.4036$ &  & $0.8029$ &  & $0.0796$ &  & $0.7277$ &  & $74.045$ &  & $347.56$ &  & $315.45$ &  & $92.871$ &  & $2076.51$ &  & $4.7788$ &  & $2.8926$ &  & $0.9538$ &  & $2.8751$ \\
 &  & BH &  & $0.4040$ &  & $0.8028$ &  & $0.0797$ &  & $0.7283$ &  & $74.358$ &  & $346.86$ &  & $315.46$ &  & $92.989$ &  & $2076.46$ &  & $4.4261$ &  & $2.5700$ &  & $0.8988$ &  & $2.6316$ \\
\hline
\multirow{2}{*}{S9} &  & RAR &  & $0.2750$ &  & $0.6430$ &  & $0.0982$ &  & $0.4518$ &  & $82.682$ &  & $150.58$ &  & $156,71$ &  & $52.259$ &  & $2028.27$ &  & $1.4582$ &  & $1.1680$ &  & $0.4809$ &  & $1.0357$ \\
 &  & BH &  & $0.2745$ &  & $0.6425$ &  & $0.0981$ &  & $0.4509$ &  & $82.532$ &  & $150.43$ &  & $156.70$ &  & $52.081$ &  & $2028.31$ &  & $1.2234$ &  & $1.4709$ &  & $0.4834$ &  & $1.0592$ \\
\hline
\multirow{2}{*}{S12} &  & RAR &  & $0.2986$ &  & $0.8812$ &  & $0.0334$ &  & $0.5638$ &  & $33.374$ &  & $318.09$ &  & $230.10$ &  & $59.232$ &  & $2054.98$ &  & $1.1906$ &  & $1.6396$ &  & $0.1657$ &  & $0.9986$ \\
 &  & BH &  & $0.2988$ &  & $0.8883$ &  & $0.0334$ &  & $0.5642$ &  & $33.520$ &  & $317.98$ &  & $230.37$ &  & $59.145$ &  & $2054.90$ &  & $1.4464$ &  & $1.5421$ &  & $0.1640$ &  & $1.0508$ \\
\hline
\multirow{2}{*}{S13} &  & RAR &  & $0.2630$ &  & $0.4275$ &  & $0.1506$ &  & $0.3754$ &  & $24.479$ &  & $245.15$ &  & $74.887$ &  & $48.856$ &  & $2004.88$ &  & $2.1403$ &  & $0.7632$ &  & $0.1324$ &  & $1.0120$ \\
 &  & BH &  & $0.2631$ &  & $0.4260$ &  & $0.1510$ &  & $0.3751$ &  & $24.479$ &  & $245.26$ &  & $74.942$ &  & $48.860$ &  & $2004.90$ &  & $2.2392$ &  & $0.7807$ &  & $0.1326$ &  & $1.0508$ \\
\hline
\multirow{2}{*}{S14} &  & RAR &  & $0.2890$ &  & $0.9564$ &  & $0.0126$ &  & $0.5654$ &  & $100.66$ &  & $337.71$ &  & $226.30$ &  & $56.422$ &  & $1999.72$ &  & $2.0007$ &  & $1.6106$ &  & $1.1548$ &  & $1.5887$ \\
 &  & BH &  & $0.2889$ &  & $0.9564$ &  & $0.0126$ &  & $0.5652$ &  & $100.40$ &  & $336.74$ &  & $226.46$ &  & $56.232$ &  & $1999.79$ &  & $1.4896$ &  & $1.9268$ &  & $1.2899$ &  & $1.5688$ \\
\hline
\multirow{2}{*}{S17} &  & RAR &  & $0.3563$ &  & $0.3973$ &  & $0.2148$ &  & $0.4979$ &  & $96.624$ &  & $324.19$ &  & $191.63$ &  & $77.315$ &  & $2067.95$ &  & $1.9937$ &  & $1.3863$ &  & $0.1222$ &  & $1.1674$ \\
 &  & BH &  & $0.3568$ &  & $0.3974$ &  & $0.2150$ &  & $0.4986$ &  & $96.636$ &  & $324.07$ &  & $191.49$ &  & $77.180$ &  & $2067.78$ &  & $1.9099$ &  & $1.3733$ &  & $0.1218$ &  & $1.1350$ \\
\hline
\multirow{2}{*}{S18} &  & RAR &  & $0.2383$ &  & $0.4716$ &  & $0.1259$ &  & $0.3507$ &  & $110.53$ &  & $350.61$ &  & $49.130$ &  & $42.297$ &  & $2036.18$ &  & $1.0372$ &  & $1.3511$ &  & $1.0739$ &  & $1.1541$ \\
 &  & BH &  & $0.2384$ &  & $0.4715$ &  & $0.1260$ &  & $0.3508$ &  & $110.53$ &  & $349.87$ &  & $49.174$ &  & $42.154$ &  & $2036.00$ &  & $1.0055$ &  & $2.6843$ &  & $1.0648$ &  & $1.5849$ \\
\hline
\multirow{2}{*}{S19} &  & RAR &  & $0.5190$ &  & $0.7510$ &  & $0.1292$ &  & $0.9088$ &  & $71.910$ &  & $155.20$ &  & $344.66$ &  & $135.46$ &  & $2005.48$ &  & $1.2719$ &  & $2.4830$ &  & $1.0759$ &  & $1.6103$ \\
 &  & BH &  & $0.5191$ &  & $0.7506$ &  & $0.1295$ &  & $0.9087$ &  & $72.034$ &  & $155.11$ &  & $344.73$ &  & $135.43$ &  & $2005.40$ &  & $1.8951$ &  & $3.1838$ &  & $0.8359$ &  & $1.9716$ \\
\hline
\multirow{2}{*}{S21} &  & RAR &  & $0.2192$ &  & $0.7622$ &  & $0.0521$ &  & $0.3863$ &  & $58.622$ &  & $166.23$ &  & $259.65$ &  & $37.210$ &  & $2027.64$ &  & $1.2499$ &  & $4.0652$ &  & $0.2691$ &  & $1.8614$ \\
 &  & BH &  & $0.2185$ &  & $0.7629$ &  & $0.0518$ &  & $0.3852$ &  & $58.630$ &  & $165.64$ &  & $259.67$ &  & $36.984$ &  & $2027.29$ &  & $1.7393$ &  & $3.7953$ &  & $0.2540$ &  & $1.9296$ \\
\hline
\multirow{2}{*}{S24} &  & RAR &  & $0.9467$ &  & $0.8908$ &  & $0.1034$ &  & $1.7900$ &  & $103.53$ &  & $289.93$ &  & $7.9969$ &  & $335.26$ &  & $2024.69$ &  & $1.6161$ &  & $3.6132$ &  & $0.1194$ &  & $1.7829$ \\
 &  & BH &  & $0.9463$ &  & $0.8907$ &  & $0.1034$ &  & $1.7892$ &  & $103.53$ &  & $289.93$ &  & $7.9990$ &  & $333.35$ &  & $2024.77$ &  & $1.2295$ &  & $3.8249$ &  & $0.1303$ &  & $1.7282$ \\
\hline
\multirow{2}{*}{S31} &  & RAR &  & $0.4472$ &  & $0.5510$ &  & $0.2008$ &  & $0.6936$ &  & $109.09$ &  & $308.04$ &  & $137.20$ &  & $108.68$ &  & $2017.98$ &  & $2.2761$ &  & $1.3093$ &  & $1.5168$ &  & $1.7007$ \\
 &  & BH &  & $0.4479$ &  & $0.5508$ &  & $0.2012$ &  & $0.6946$ &  & $108.93$ &  & $307.93$ &  & $137.19$ &  & $108.56$ &  & $2017.94$ &  & $2.7348$ &  & $1.2618$ &  & $1.5448$ &  & $1.8472$ \\
\hline
\multirow{2}{*}{S38} &  & RAR &  & $0.1408$ &  & $0.8175$ &  & $0.0257$ &  & $0.2559$ &  & $170.98$ &  & $18.053$ &  & $99.694$ &  & $19.182$ &  & $2003.26$ &  & $1.3141$ &  & $2.6440$ &  & $0.4762$ &  & $1.4781$ \\
 &  & BH &  & $0.1411$ &  & $0.8195$ &  & $0.0255$ &  & $0.2567$ &  & $170.98$ &  & $18.215$ &  & $99.761$ &  & $19.195$ &  & $2003.31$ &  & $1.3480$ &  & $2.5486$ &  & $0.4758$ &  & $1.4575$ \\
\hline
\multirow{2}{*}{S54} &  & RAR &  & $1.1985$ &  & $0.8921$ &  & $0.1293$ &  & $2.2676$ &  & $62.242$ &  & $140.76$ &  & $288.44$ &  & $478.38$ &  & $2004.30$ &  & $1.1884$ &  & $1.5459$ &  & $0.2956$ &  & $1.0099$ \\
 &  & BH &  & $1.1986$ &  & $0.8927$ &  & $0.1287$ &  & $2.2685$ &  & $62.188$ &  & $140.79$ &  & $288.44$ &  & $475.18$ &  & $2004.38$ &  & $1.5915$ &  & $1.1222$ &  & $0.2922$ &  & $1.0020$ \\
\hline
\multirow{2}{*}{S55} &  & RAR &  & $0.1082$ &  & $0.7206$ &  & $0.0302$ &  & $0.1861$ &  & $149.93$ &  & $331.33$ &  & $325.45$ &  & $12.905$ &  & $2009.29$ &  & $0.4437$ &  & $2.0672$ &  & $3.1088$ &  & $1.8732$ \\
 &  & BH &  & $0.1083$ &  & $0.7204$ &  & $0.0303$ &  & $0.1863$ &  & $149.94$ &  & $331.44$ &  & $325.48$ &  & $12.908$ &  & $2009.30$ &  & $0.4504$ &  & $1.9576$ &  & $3.1099$ &  & $1.8393$ \\
 \hline
 &  &  &  &  &  &  &  &  &  &  &  &  &  &  &  &  &  &  &  &  &  &  &  &  &  &  &  &  
\end{tabular}%
}
\begin{tablenotes}%
\item $a$: semi-major axis of the orbit; $e$: eccentricity; $r_p$: distance to pericenter; $r_a$: distance to apocenter; $i$: inclination; $\omega$: argument of pericenter; $\Omega$: position angle of the ascending node; $P$: orbital period; $t_p$: epoch of pericenter passage; $\bar{\chi}^2_X$: reduced-$\chi^2$ for $X$ position; $\bar{\chi}^2_Y$: reduced-$\chi^2$ for $Y$ position; $\bar{\chi}^2_z$: reduced-$\chi^2$ for line-of-sight radial velocity; $\langle \bar{\chi}^2\rangle = \frac{1}{3}\left( \bar{\chi}^2_{X} + \bar{\chi}^2_{Y} + \bar{\chi}^2_{z} \right)$. 

\item \textbf{Note}: We refer to \citet{2020A&A...641A..34B} for details on the definition of the parameters and on the fitting procedure.
\end{tablenotes}
\end{center}
\end{table}
\end{landscape}

\bsp	
\label{lastpage}
\end{document}